# Raman signatures of pressure induced electronic topological and structural transitions in $Bi_2Te_3$

Gopal K. Pradhan, Achintya Bera, Pradeep Kumar, D V S Muthu and A K Sood[a)]

Department of Physics, Indian Institute of Science, Bangalore - 560012, India

## Abstract

We report Raman signatures of electronic topological transition (ETT) at 3.6 GPa and rhombohedral (α-$Bi_2Te_3$) to monoclinic (β-$Bi_2Te_3$) structural transition at ~ 8 GPa. At the onset of ETT, a new Raman mode appears near 107 $cm^{-1}$ which is dispersionless with pressure. The structural transition at ~ 8 GPa is marked by a change in pressure derivative of $A_{1g}$ and $E_g$ mode frequencies as well as by appearance of new modes near 115 $cm^{-1}$ and 135 $cm^{-1}$. The mode Grüneisen parameters are determined in both the α and β-phases.

[a)]Electronic mail- asood@physics.iisc.ernet.in

1. **Introduction**

Bismuth telluride ($Bi_2Te_3$) has been shown to be one of the simplest three dimensional (3-D) topological insulators, [1,2] an extraordinary thermoelectric compound at ambient temperature [3-5] and a possible topological superconductor [6]. More recently, atomically-thin layers have also been isolated by mechanical exfoliation from bulk $Bi_2Te_3$ in a "graphene-like" fashion for enhanced thermoelectric properties [7]. All these have made $Bi_2Te_3$ (BT) a subject of intense investigation both in basic and applied research. High-pressure (HP) studies on BT have unveiled many interesting phenomena such as giant improvement of thermoelectric power factor at 1 GPa [3], superconductivity [6,8] and reconstruction of Fermi surface topology giving rise to an electronic topological transition (ETT) or Lifschitz transition at ~3 GPa [9,10]. The pressure-induced ETT is an isostructural transition without any volume discontinuity and the Wyckoff positions of the atoms are not modified during the transition. The pressure dependence of the structural parameters related to interlayer weak van der Waals interactions is insensitive to the modification of the Fermi surface whereas structural parameters related to the strong covalent-ionic (ionocovalent) bonds along the layer planes are significantly affected by the ETT with a 25% increase in the bulk modulus and a 68% decrease in its pressure derivative [9]. This transition related to the modification of the compressibility only in the layer planes has been shown to influence the thermoelectric properties strongly [5]. It has also been shown that the pressure coefficient of the band gap is −20 meV/GPa below and −60 meV/GPa above the ETT [5]. HP x-ray diffraction (XRD) experiments [10,11] have demonstrated that BT (α-phase) transforms into HP phases II (β- $Bi_2Te_3$) and III (γ- $Bi_2Te_3$) at 8.0 and 14.0 GPa, respectively. More recent and detailed XRD studies [12,13] established that above 14.4 GPa, BT evolves into Bi-Te substitutional alloy (phases IV) by adopting a body-centered cubic (bcc) disordered structure stable at least up to 52.1 GPa. While phase II and III adopt monoclinic seven-fold BiTe$_7$ (*C2/m*) and eight-fold BiTe$_8$ (*C2/c*) structures, respectively, phase IV is cubic ($Im\bar{3}m$) [12].

High pressure Raman spectroscopy known to be a powerful probe for structural phase transition has not been reported for BT except one study in the low pressure (< 1 GPa) regime [14]. It is pertinent to ask if there are any vibrational signatures of ETT under pressure. In this letter, we report HP Raman studies on Bi$_2$Te$_3$ single crystals up to 14 GPa to investigate the ETT and the structural transitions i.e. α → β at ~8 GPa and β → γ at ~ 14 GPa. On completion of this work, we became aware of a recent high pressure Raman, optical absorption and reflection study of Bi$_2$Te$_3$ [20] upto 23GPa. We will bring out similarities and differences between our study and Vilaplana et al. [20].

## 2. Experimental details

Thin platelets (~30–40µm thick) cleaved from BT single crystals were placed together with a ruby chip into a stainless steel gasket inserted between the diamonds of a Mao/Bell-type diamond anvil cell (DAC). Methanol-ethanol (4:1) mixture was used as the pressure transmitting medium, the pressure being determined via the ruby fluorescence scale [15]. Unpolarized Raman spectra were recorded in backscattering geometry using the 514.5 nm excitation from an Ar$^+$ ion laser (Coherent Innova 300). The spectra were collected by a DILOR XY Raman spectrometer coupled to a liquid nitrogen cooled charged coupled device, (CCD 3000 Jobin Yvon-SPEX). The pixel resolution is 0.85 cm$^{-1}$ and the instrumental resolution is better than ~5 cm$^{-1}$. After each Raman measurement, calibration spectra of a Ne lamp were recorded to correct for small drifts, if any, in the energy calibration of the spectrometer. Laser power (< 5 mW) was held low enough to avoid heating of the sample. The peak positions were determined by fitting Lorentzian line shapes with an appropriate background.

## 3. Results and discussion

BT has a layered structure having rhombohedral symmetry ($R\bar{3}m$) having lattice parameters $a$ = 4.383 Å and $c$ = 30.38 Å (in the hexagonal setting) [16]. The atoms are arranged in planes perpendicular to the *c*-axis and form layers of five planes in the sequence [Te(2)-Bi-Te(1)-Bi-Te(2)] where the Bi and the Te(2) atoms occupy the 6*c* Wyckoff sites and the Te(1) atoms the 3*a* sites. The bonds within the layer are ionocovalent bonds, and the layers are bonded by van der Waals interaction. BT has five atoms in its unit cell giving rise to 12 optical phonons characterized by $2A_{2u} + 2E_u + 2A_{1g} + 2E_g$ symmetry with $E_g$ and $A_{1g}$ being Raman active, where the atoms vibrate in-plane ($E_g$) and out-of-plane ($A_{1g}$) [17]. Raman spectra at ambient pressure (outside the DAC) show three Raman modes: $A_{1g}^1$ at ~62 cm$^{-1}$, $E_g^2$ at ~102 cm$^{-1}$, and $A_{1g}^2$ at ~135 cm$^{-1}$, in good agreement with previous reports in bulk and thin films of BT [17-19]. However, due to low signal to noise ratio and high scattering background inside the DAC, only $E_g^2$ and $A_{1g}^2$ modes could be observed and followed in our HP experiments.

Fig. 1 shows the pressure evolution of the Raman spectra at a few representative pressures in the increasing pressure run. It can be seen (marked by arrows in Fig. 1) that a new mode M1 starts appearing at 3.6 GPa and M2 and M3 modes beyond 8 GPa. Beyond 12 GPa, the peaks become very broad and the signal becomes extremely weak and we could not follow any mode beyond 14 GPa. Upon decreasing the pressure, Raman modes start appearing only at ~7 GPa (Fig. 2), showing a considerable hysteresis in the pressure induced changes. Raman spectrum of the pressure recycled sample at 0.6 GPa is same as that of the starting sample. Fig. 3 shows the pressure-dependence of various Raman modes. The solid lines are linear fits to the data using $\omega_P = \omega_0 + \left(\dfrac{d\omega}{dP}\right)P$ and the values of $\omega_0$ and the pressure derivative $\frac{d\omega}{dP}$ are given in Table 1. The following observations can be made: (i) beyond 3.6 GPa, a new Raman mode (M1) with negligible pressure dependence is seen at 107 cm$^{-1}$ and the onset of this mode coincides with ETT [9,10] at ~ 3 GPa; (ii) At ~ 8 GPa, two new modes at ~ 118 cm$^{-1}$ (M2) and 139 cm$^{-1}$ (M3) appear; and (iii) The pressure dependence of the $E_g^2$ and $A_{1g}^2$ modes changes after ~8 GPa. We associate these last two observations with the phase transition from rhombohedral phase (phase I termed as α-Bi$_2$Te$_3$) to monoclinic phase with symmetry *C2/m* (termed as phase II or β- Bi$_2$Te$_3$) [8,12]. This symmetry lowering gives rise to the new modes because the sevenfold (BiTe$_7$) Bi$_2$Te$_3$ monoclinic structure has 15 Raman active modes at the Γ-point of the Brillouin zone, represented by 10A$_g$+5B$_g$. We could not observe the additional modes. Above 14 GPa, we could not observe any Raman mode even after acquiring the spectra with very long exposure times (30 minutes). This marks the transition to phase III. We believe that the highly metallic nature [8] of phase III gives rise to an extremely small skin depth thus significantly reducing the scattering volume and hence the intensity. The broadening of the modes beyond 10 GPa can be partly due to non-hydrostatic nature of the pressure medium. The behavior on decompression is interesting as shown in Fig. 3. After releasing the pressure from 14.2 GPa, the Raman modes are seen only at 7 GPa and below. The pressure dependence of the $E_g^2$ mode is same in both increasing and decreasing pressure runs. This should be contrasted with the different behavior of the $A_{1g}^2$ mode frequencies (see Fig. 3). The extra mode at 107 cm$^{-1}$ marking the ETT at ~ 3.6 GPa in the increasing pressure run is not seen below 2.6 GPa in the decreasing pressure run. Now coming to the origin of this new mode M1, we can only offer a plausible qualitative explanation. The $E_g^2$ mode involves the vibration of Bi and Te (2) in the basal plane and ETT only modifies the compressibility parallel to the layers. Even though the overall symmetry of the crystal is not observed to be affected in XRD due to ETT [9], we propose that some local distortions can occur in the basal plane, lifting the degeneracy of the $E_g^2$ mode and hence appearance of mode M1 beyond 3.6 GPa. More theoretical work is needed to understand the appearance and negligible pressure dependence of this mode. Individual mode Grüneisen parameters ($\gamma_i$) can be calculated using the relation $\gamma_i = \frac{B_0}{\omega_i}\left(\frac{d\omega}{dP}\right)$, where $B_0$ is the bulk modulus. In the 0-8 GPa pressure regime, there exist a range of $B_0$ values [9-11] which are close to each other. We take the average of their $B_0$ values which is 36.1 GPa for phase I, whereas we take $B_0$ to be 112 GPa (Ref. 10) for phase II. Table 1 lists the pressure coefficients ($d\omega/dP$) and the Grüneisen parameters. For comparison, pressure coefficients from Ref. 17 (P < 1 GPa) for $E_g^2$ and $A_{1g}^1$ mode are 3.4 and 4 cm$^{-1}$/GPa, respectively. It can be seen that $A_{1g}^1$ mode has higher pressure coefficients than the other two modes. This difference can be qualitatively understood taking into account the eigenvectors [17]. It can be recalled that in the $A_{1g}^1$ mode, Bi and Te(2) atoms are vibrating in-phase (along *c*- axis) while in both $A_{1g}^2$ and $E_g^2$ modes, Bi and Te(2) atoms vibrate out of phase with the displacement along the *c*- and *a*- axis, respectively. The Te(2) atoms in adjacent sandwiches vibrate out of phase. Thus, in the low-frequency $A_{1g}^1$ vibration, the pressure variation of frequency is influenced by the weak inter-sandwich Te(2)-Te(2) van der Waal interactions and hence the higher pressure coefficient. It can be noticed from Table 1 that the pressure derivative of the $E_g^2$ mode is about 20% more than the $A_{1g}^2$ mode. Taking the bulk modulus $B_0$ = 36.1 GPa, $\gamma_i$ for both the modes in phase I are 1.2 ($E_g^2$) and 0.75 ($A_{1g}^2$). However, in the HP phase (phase II), $\gamma_i$ for all the modes are very similar which is usually the case of 3-D network solids. At elevated pressure, anisotropy in the structure where Bi is seven (phase II) or eight (phase III) coordinated with Te through monoclinic distortions [12].

We now compare our results with the recent work by Vilaplana et al [20] and Polian et al [9]. In the work of Vilaplana et al [20], the pressure coefficients of the two modes in α-$Bi_2Te_3$ are in good agreement with our results. However, unlike our study, they do not find a new mode to mark the onset of ETT at 3.6 GPa. We do not know the reason for this. We speculate that the appearance of the new mode M1 in our studies may be related to the change in the compressibility parallel to the layers, as shown in ref. [9]. In ref. [20], the pressure dependence of the FWHM of the Raman allowed modes $A_{1g}^1$, $E_g^2$ and $A_{1g}^2$ shows a change in slope at ~ 4 GPa. This was interpreted to suggest that ETT results in different structural compressibility in directions parallel and perpendicular to the layers. This is in disagreement with the conclusions of Polian *et al* [9] derived from high pressure x-ray diffraction studies. However, our FWHM data did not reveal such changes. An additional feature in our work is the Raman spectra on release of pressure (Fig.3). We show that there is considerable hysteresis, especially in the pressure dependence of the $A_{1g}^2$ mode. Further, the additional mode near 107 cm$^{-1}$ is absent in the pressure released $Bi_2Te_3$ below 2.6 GPa.

## 4. Conclusion

In summary, high-pressure Raman experiments on $Bi_2Te_3$ have revealed a new Raman mode at ~3.6 GPa coinciding with the onset of the ETT. We suggest that it is related to local distortions in the basal plane following the ETT. A structural transition to monoclinic phase is evidenced at ~8 GPa in agreement with recent XRD measurements. The metallization process is gradual completing at ~ 14 GPa. The mode Grüneisen parameters also suggest that as compared to the parent layered structure with ionocovalent and van der Waal bonding, high pressure phase II has a network structure with dominantly one type of bonding. We hope that our experiments will motivate first-principles calculations of phonons and electronic structure at high-pressure which will also lead to an understanding of pressure enhanced thermoelectric power factor.

**Acknowledgements**

AKS thanks the DST, India for funding. AB and PK thank CSIR for a fellowship.

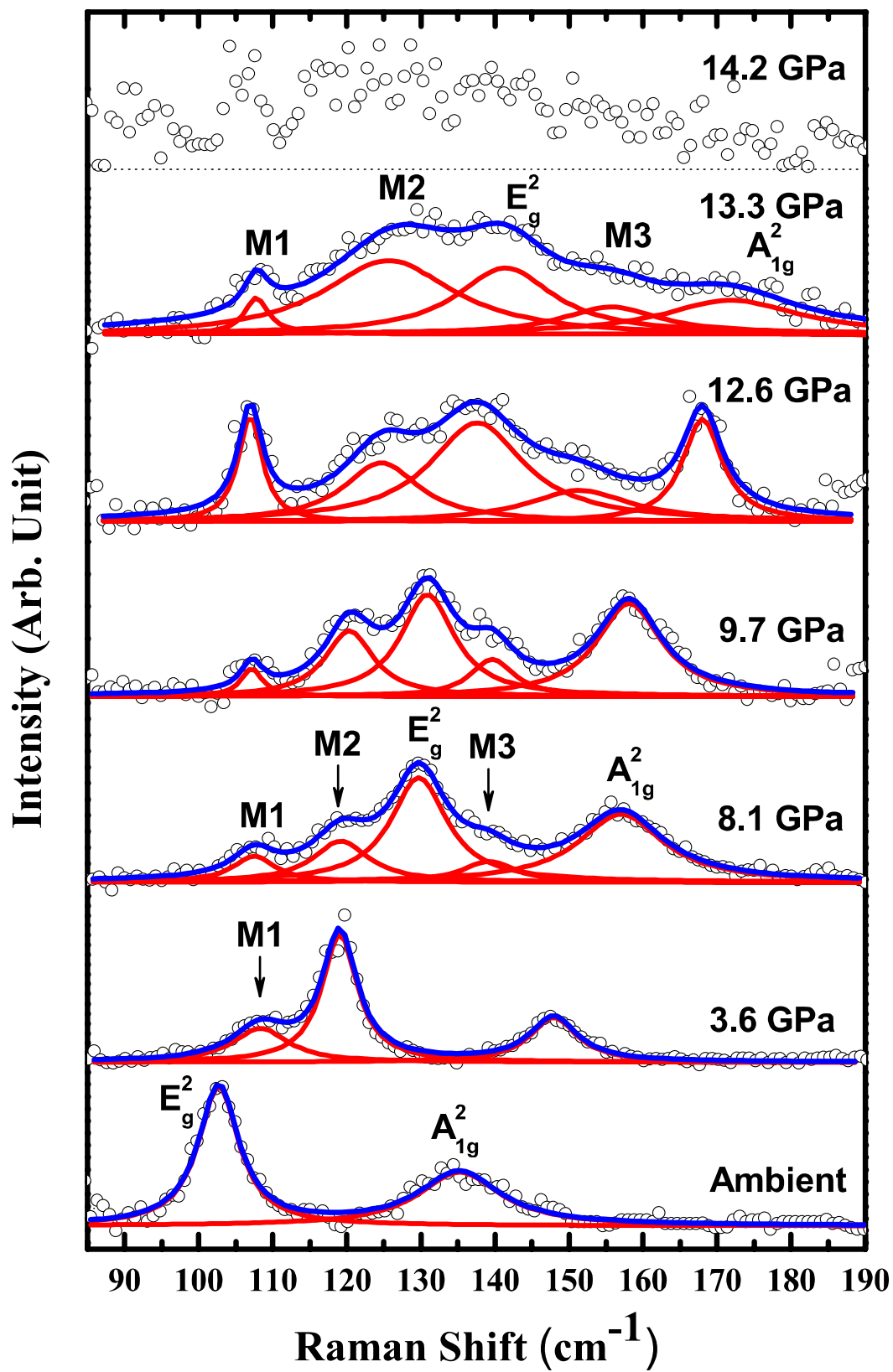

**Figure 1** (color online) – Pressure evolution of Raman spectra. The solid lines are Lorentzian fits to the experimemtal data points (open circles). Appearance of new peaks (M1, M2 and M3) is indicated by arrows. The dashed horizontal line at the top shows the base line.

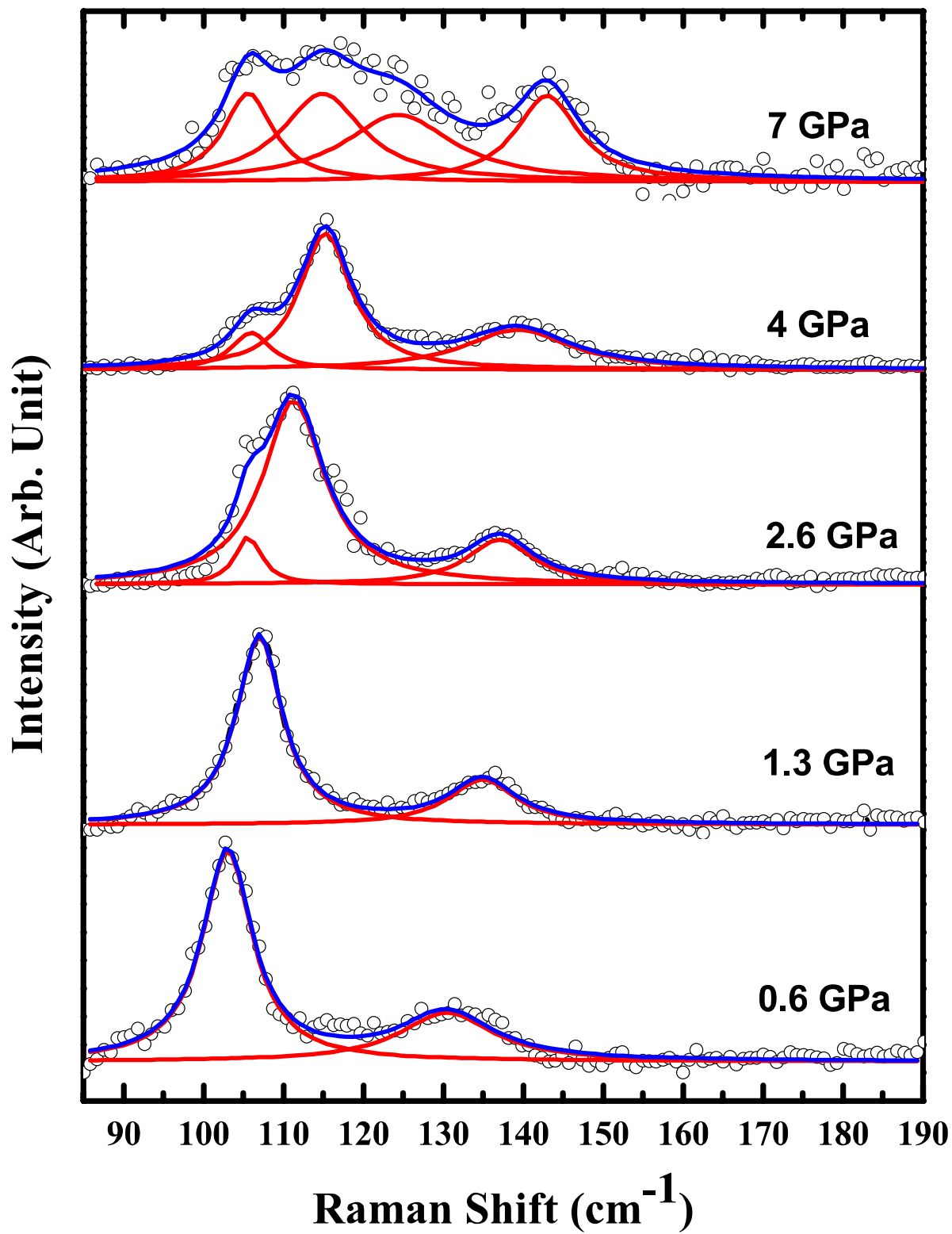

**Figure 2** (color online) - Pressure Evolution of Raman spectra in the return pressure cycle. The solid lines are Lorentzian fits to the experimemtal data points (open circles).

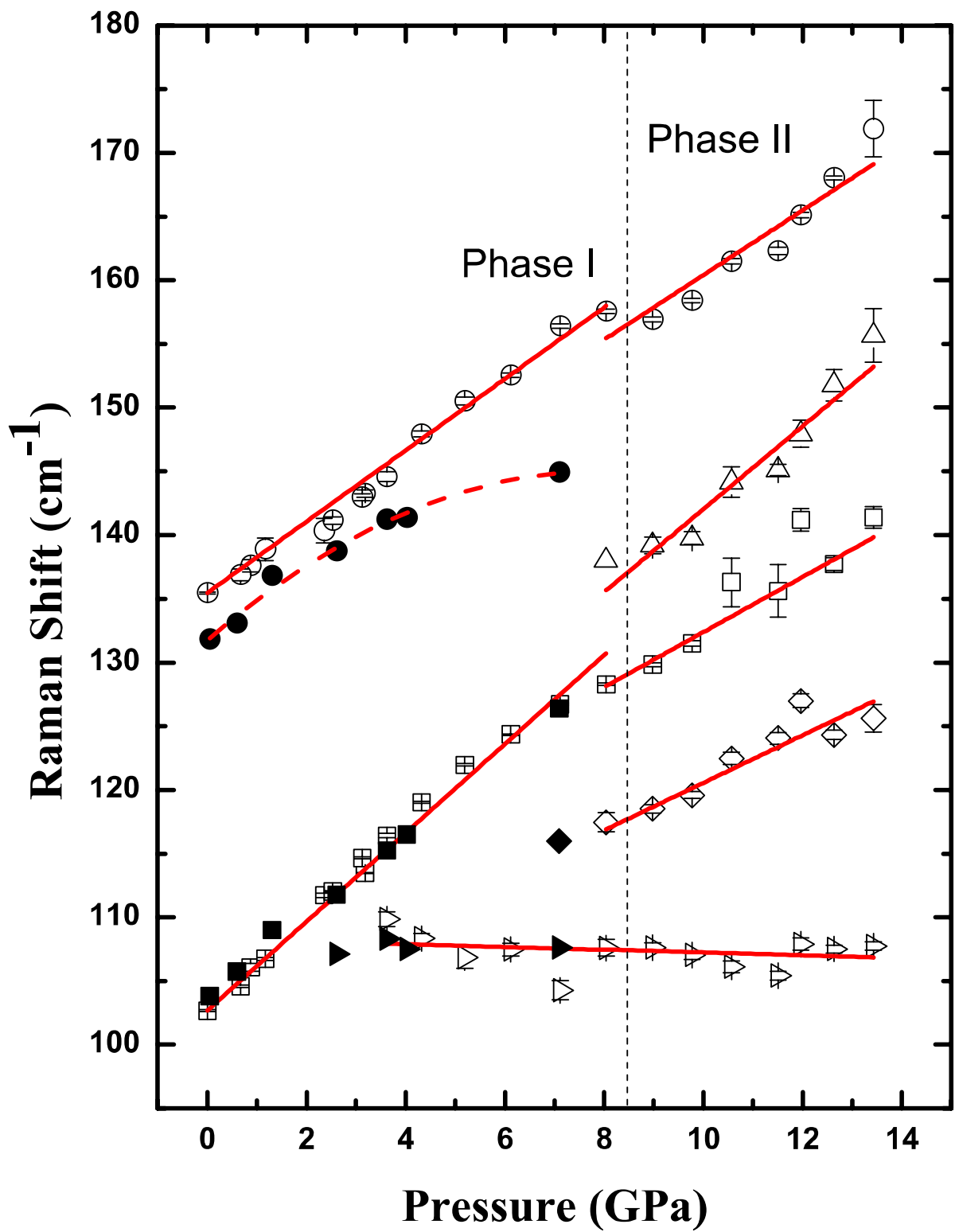

**Figure 3** (color online) - Frequency versus pressure plot for various Raman modes. The solid lines are linear fits to the observed frequencies (open symbols). Error bars (obtained from the fitting procedure) are also shown. The filled symbols represent the observed frequencies in the return pressure run, the dashed line being a guide to the eye. The vertical (dashed) line indicates the phase transition pressure.

**Table 1 -** Pressure coefficient (dω/dP) and mode Grüneisen parameters ($\gamma_i$) for various modes

| Phase | Mode frequeny ($\omega_0$) (cm$^{-1}$) | dω/dP (cm$^{-1}$GPa$^{-1}$) | $\gamma_i$ |
|---|---|---|---|
| **Phase I** | 102.6 ±0.1 ($E_g^2$) | 3.5 ±0.05 | 1.2 |
| | 135.4 ±0.3 ($A_{1g}^2$) | 2.8 ±0.04 | 0.75 |
| | 107.3 ±0.6 | -0.1 ±0.1 | -0.04 |
| **Phase II** | 129.8±0.2 (107.9[a]) | 2.5 ±0.2 | 2.5 |
| | 156.9 ±0.2, (131[a]) | 2.9± 0.1 | 2.4 |
| | 118.5±0.3, (101.3[a]) | 1.9 ±0.2 | 2.1 |
| | 139.2 ±0.6, (115.3[a]) | 2.6 ±0.3 | 2.6 |

[a] This is the extrapolated P = 0 frequency for the modes seen in the high pressure phase II.